# DFT based comparative analysis of physical properties of binary metallic diborides XB$_2$ (X = Cr, Mo and W)

Razu Ahmed[1,2], Md. Sohel Rana[1], Md. Sajidul Islam[1], S. H. Naqib[1*]

[1]Department of Physics, University of Rajshahi, Rajshahi-6205, Bangladesh

[2]Department of Physics, Bangladesh Army University of Engineering and Technology, Qadirabad-6431, Bangladesh

*Corresponding author, Email: salehnaqib@yahoo.com

**Abstract**

Transition-metal borides (TMBs) have long attracted attention of the researchers because of their unique mechanical and electrical properties including superconductivity. We have explored the structural, mechanical, electronic, optical, and some thermophysical properties of XB$_2$ (X = Cr, Mo and W) binary metallic diborides in detail employing density functional theory based first-principles method. Many of the physical properties, including direction-dependent mechanical properties, optical properties, and thermo-mechanical properties are being investigated for the first time. According to this study, XB$_2$ (X = Cr, Mo and W) compounds exhibit reasonably good machinability, delocalized metallic bonding, and high Vickers hardness with high Debye temperature. CrB$_2$ and MoB$_2$ exhibit brittleness whereas WB$_2$ show ductility. The mechanical stability of XB$_2$ (X = Cr, Mo and W) compounds has been confirmed. Based on the calculated bulk modulus, Young's modulus and shear modulus of XB$_2$ (X = Cr, Mo and W) compounds can be regarded as potential candidates of ultra-incompressible and hard materials. The metallic nature of XB$_2$ (X = Cr, Mo and W) compounds is confirmed via electronic band structures with a high electronic energy density of states at the Fermi level. The optical parameters exhibit excellent agreement with the electrical properties. All optical constants show a considerable degree of anisotropy. The reflectivity spectra reveal that XB$_2$ (X = Cr, Mo and W) compounds are good reflectors in the infrared and near-visible regions. The compounds under study have a high Debye temperature, melting temperature, lattice thermal conductivity, and minimum phonon thermal conductivity, which corresponds well to their elastic and bonding properties. The extremely high melting temperature of WB$_2$ indicates that WB$_2$ is a promising material for high-temperature applications. The superconducting state parameters of XB$_2$ (X = Cr, Mo and W) compounds are also investigated.

**Keywords:** Density functional theory; Metallic diborides; Thermo-mechanical properties; Optoelectronic properties; Superconductivity

## 1. Introduction

Transition-metal borides (TMBs) have a wide range of practical uses due to their unique physical properties such as large bulk modulus, high hardness, ultrahigh melting points, favorable thermal stability, strong resistance to oxidation, excellent electric transport properties, and superconductivity at high temperature [1–3]. These have great potential for application in harsh environments such as those encountered by hypersonic flight, scramjet and rocket propulsion, atmospheric re-entry, hard coatings for electromechanical systems, armor, and cutting tools [1–5].

In addition to the fascinating chemical properties already described, metal diborides also take on distinct crystal structures based on the $R_M/R_B$ ratio, where $R_M$ and $R_B$ are atomic radii of metallic and boron atoms, respectively. The most representative category of TMBs is generated by the layered hexagonal diborides with $AlB_2$-type structure which features a graphitelike boron layer alternated with a close-packed metal layer [6,7]. Strong B–B bonds in the boron layer provide structural stability. The range of the $R_M/R_B$ ratio for these stable $AlB_2$-like TMB phases is 1.14 to 2.06 [8,9]. Superhard $ReB_2$ has a hexagonal structure with a unique zigzag network generated by the strong Re–B and B–B bonds [6,10] which contributes significantly to its high hardness. However, subsequent experiments identified that the structure of $XB_2$ (X = Cr, Mo and W) should have an $AlB_2$-type structure (P6/*mmm*, No. 191) [11–13].

The surprising discovery of superconductivity in magnesium diboride ($MgB_2$) with superconducting transition temperature $T_c$ ~ 39 K has provided a fresh and strong impetus for research into this family of materials [14 - 18]. Superconductivity has also been observed in $OsB_2$ ($T_c$ = 2.1 K) and $RuB_2$ ($T_c$ = 1.6 K) [16,17]. The electron-phonon interaction is the driving force behind the observed superconductivity in each of these compounds, which show the wide range of material types and applications that include B-B bonds and metals. According to theoretical calculations [19,20], superconductivity of $MoB_2$ in the $MgB_2$ type structure at high pressure still falls within the traditional BCS-type. Furthermore, the compound $CrB_2$ demonstrates superconductivity at a transition temperature of up to 7 K when external pressure is applied, but superconductivity of $CrB_2$ is detected when the antiferromagnetic transition at $T_N$ ~ 88 K is totally suppressed under ambient pressure [21].

In the present work, we employed first-principles calculations based on the density functional theory (DFT) to investigate the crystal structures of $XB_2$ (X = Cr, Mo and W) compounds. Some earlier experimental and theoretical findings on the selected materials were reported in Refs. [11–13,22,23]. However, there is still a noticeable lack of information on these materials. This lack of information reduces the viability of its possible applications. To the best of our knowledge, many important physical properties (e.g., elastic, optoelectronic, and thermo-physical properties) have yet to be studied. Additionally, the mechanical anisotropy characterized by the direction dependence of the shear modulus, Young's modulus, linear compressibility, and Poisson's ratio of $CrB_2$ and $WB_2$ compounds has not been discussed at all, except for $MoB_2$ [17].

In this study, we conducted a comparative analysis of the physical properties of $XB_2$ (X = Cr, Mo and W) compounds in order to completely grasp their structural, elastic, electronic, optical, thermo-mechanical, and superconducting state properties. The results obtained are compared with those found in previous studies where available. When selecting a system for applications involving optoelectronic devices, optical properties are crucial to understand. The behavior of a material at various temperatures may be inferred from thermo-physical properties. Mechanical anisotropy is crucial for understanding key mechanical behaviors such as crack formation and its subsequent propagation, crystal structure instability, phase transformation, and growth of plastic deformations which sometimes restrict efficient applications of materials. Additionally, this study computes bond hardness, fracture toughness, and brittleness index, which are useful for designing any part of a structure or device. a comprehensive understanding of the elastic, mechanical, thermomechanical, and optical response of these compounds is necessary to unravel the full promise of $XB_2$ (X = Cr, Mo and W) for potential applications. This constitutes the primary motivation of the present study.

The rest of the paper is organized as follows: In Section 2, we have discussed the computational scheme. Section 3 reveals the computational results and analyses. Finally, the major findings of this study are discussed and summarized in Section 4.

## 2. Computational scheme

Ground state study of the systems is performed with the Kohn-Sham density functional theory (KS-DFT) [24] using the CAmbridge Serial Total Energy Package (CASTEP) code [25]. The generalized gradient approximation (GGA) with Perdew–Burke–Ernzerhof (PBE) [26], and local density approximation (LDA) [27,28] have been employed for the exchange-correlation functionals. Optimized geometry is obtained using GGA (PBE) for $CrB_2$ and LDA for $MoB_2$ and $WB_2$. Ultrasoft Vanderbilt-type pseudopotentials [29] have been employed to model the electron-ion interactions and represent electronic wavefunctions using a plane wave basis set [30]. We treated $[3s^23p^63d^54s^1]$, $[4s^24p^64d^55s^1]$, $[5s^25p^65d^46s^2]$, and $[2s^22p^1]$ as valence electron configurations for Cr, Mo, W, and B atoms, respectively. In this paper, Monkhorst-Pack grid of $k$-point meshes with the sizes of 12 × 12 × 10 for $CrB_2$ and $WB_2$, and 21 × 21 × 18 for $MoB_2$ have been selected and the cut-off energy for the plane wave expansion is set to 550 eV. The BFGS (Broyden-Fletcher-Goldfarb-Shanno) algorithm [31] was used as the minimization algorithm, and density mixing was used to determine the electronic structure. The convergence criterion of relaxation on the lattice unit cell has been kept within a limit of $5\times10^{-6}$ eV-atom$^{-1}$ for energy, 0.01 eV Å$^{-1}$ for maximum force, 0.02 GPa for maximum stress, and $5\times10^{-4}$ Å for maximum atomic displacement.

The elastic constants, $C_{ij}$, are acquired from the stress-strain relationship [32]. The Voigt-Reuss-Hill (VRH) method [33,34] is used to estimate all the other polycrystalline elastic parameters, including the bulk modulus ($B$), shear modulus ($G$), and Young's modulus ($Y$) and Poisson ratio ($\sigma$).

The energy/frequency dependent complex dielectric function, $[\varepsilon(\omega) = \varepsilon_1(\omega) + i\varepsilon_2(\omega)]$, can be used to obtain all the optical parameters. CASTEP uses the following expression to calculate the frequency dependent imaginary part of the dielectric function:

$$\varepsilon_2(\omega) = \frac{2e^2\pi}{\Omega\varepsilon_0}\sum_{k,v,c}|\langle\psi_k^c|\hat{u}.\vec{r}|\psi_k^v\rangle|^2 \delta(E_k^c - E_k^v - E) \qquad (1)$$

where, $\Omega$ is the cell volume, $\omega$ is the angular frequency of incident electromagnetic wave (EMW), $\hat{u}$ is the unit vector giving the polarization direction of the electric field, $e$ is the electronic charge, $\psi_k^c$ and $\psi_k^v$ are the conduction and valence band wave functions at a fixed wave-vector $k$. The delta function in Equation (1) ensures energy and momentum conservation during the optical transition. The well-known Kramers-Kronig equation links the real and imaginary components of ε(ω):

$$\varepsilon_1(\omega) = 1 + \frac{2}{\pi} P \int_0^\infty \frac{\omega' \varepsilon_2(\omega') d\omega'}{(\omega'^2 - \omega^2)} \qquad (2)$$

In Equation (2), $P$ signifies the principal part of the dielectric function. All the other important optical parameters including absorption coefficient [$\alpha(\omega)$], optical conductivity [$\sigma(\omega)$], loss function [$L(\omega)$], reflectivity [$R(\omega)$], and refractive index [real, $n(\omega)$, and imaginary, $k(\omega)$, parts], are estimated from the dielectric function $\varepsilon(\omega)$. The interrelations are given below [35–39]:

$$\alpha(\omega) = \frac{4\pi k(\omega)}{\lambda} \qquad (3)$$

$$n(\omega) = \frac{1}{\sqrt{2}} [\{\varepsilon_1(\omega)^2 + \varepsilon_2(\omega)^2\}^{1/2} + \varepsilon_1(\omega)]^{1/2} \qquad (4)$$

$$k(\omega) = \frac{1}{\sqrt{2}} [\{\varepsilon_1(\omega)^2 + \varepsilon_2(\omega)^2\}^{1/2} - \varepsilon_1(\omega)]^{1/2} \qquad (5)$$

$$R(\omega) = \left|\frac{\tilde{n} - 1}{\tilde{n} + 1}\right| = \frac{(n-1)^2 + k^2}{(n+1)^2 + k^2} \qquad (6)$$

$$\sigma(\omega) = \frac{2 W_{cv} \hbar \omega}{\vec{E}_0^{\,2}} \qquad (7)$$

$$L(\omega) = Im\left(-\frac{1}{\varepsilon(\omega)}\right) \qquad (8)$$

In Equation (7), $W_{cv}$ is the photon induced optical transition probability per unit time.

## 3. Results and discussions
### 3.1 Structural

The compound $XB_2$ (X = Cr, Mo and W) conforms to an $AlB_2$-type layered structure with a space group of P6/*mmm* (No. 191) [11–13,15]. Figure 1 shows the schematic crystal structure of $XB_2$ (X = Cr, Mo and W). It is a simple hexagonal lattice of close-packed metal layers alternating with graphite-like B layers perpendicular to the c direction [15,40]. With three closest neighbor B atoms in each plane, the boron atoms are positioned at the corners of a hexagon. The metal atoms are found in the center of the B hexagons, midway between adjacent boron layers. One metal and two atoms of boron make up the primitive cell. There is one metal and one B in nonequivalent atomic positions of X (0, 0, 0) and B (1/3, 2/3, 1/2) [11–13,15].

The optimized structural parameters of $XB_2$ (X = Cr, Mo and W) are shown in Table 1 along with the experimental and theoretical results [11–13,22] for comparison. The computed lattice parameters are in excellent agreement with the experimental results. The calculated ratios c/a for each of the $XB_2$ phases is also presented in Table 1. Pearson's criteria states [15,41] that for stable $AlB_2$-like phases, the ratio c/a should fall between 0.59 and 1.2. Thus, from the data obtained it may be concluded that the $XB_2$ (X = Cr, Mo and W) for which c/a < 1.2, should be stable, whereas $WB_2$ (c/a = 1.12) is close to the upper limit of stability.

.

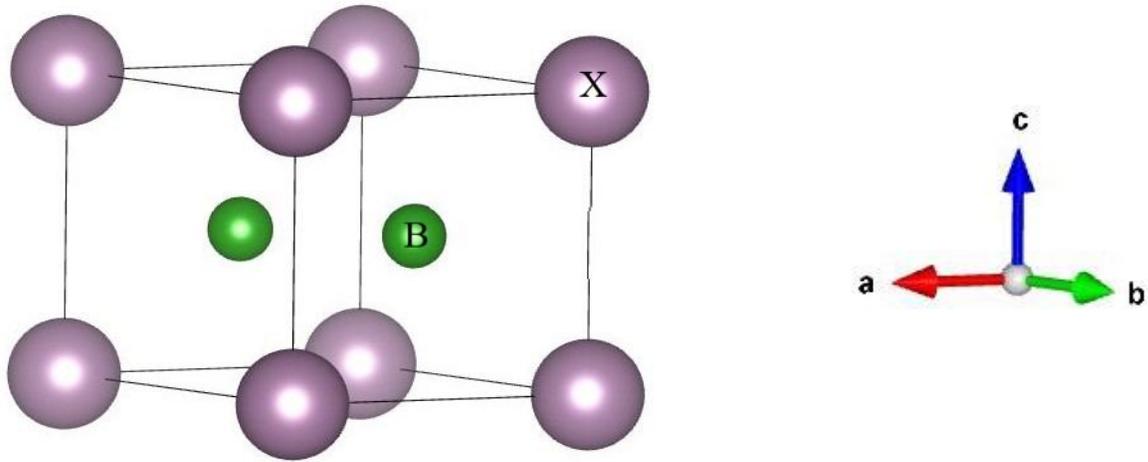

**Figure 1.** Schematic crystal structure of $XB_2$ (X = Cr, Mo and W). Crystallographic axes are shown on the right.

Table 1: Calculated lattice constants a, b and c (all in Å), equilibrium volume $V_o$ (Å$^3$), number of formula units in the unit cell Z and total number of atoms in the unit cell N of $XB_2$ (X = Cr, Mo and W) compounds.

| Compound | a | b | c | c/a | $V_o$ | Z | N | Reference |
|---|---|---|---|---|---|---|---|---|
| CrB$_2$ | 2.97 | 2.97 | 2.94 | 0.99 | 22.49 | 1 | 3 | This work |
|  | 2.98 | 2.98 | 3.07 | - | 23.50 | 1 | 3 | [11] Expt. |
| MoB$_2$ | 2.99 | 2.99 | 3.28 | 1.09 | 25.38 | 1 | 3 | This work |
|  | 3.05 | 3.05 | 3.07 | - | 24.67 | 1 | 3 | [12] Expt. |
|  | 3.01 | 3.01 | 3.31 | 1.10 | - | - | - | [15] Theo. |
| WB$_2$ | 2.97 | 2.97 | 3.32 | 1.12 | 25.43 | 1 | 3 | This work |
|  | 3.02 | 3.02 | 3.05 | - | 24.09 | 1 | 3 | [13] Expt. |
|  | 3.05 | 3.05 | 3.31 | - | 26.69 | 1 | 3 | [22] Theo. |

### 3.2 Elastic properties

In order to predict a material's response to external stresses, it is necessary to comprehend the majority of its solid-state properties, including brittleness, ductility, stiffness, structural stability, normal modes of oscillations, elastic wave propagation, and anisotropy. The values of five independent elastic constants ($C_{ij}$, namely $C_{11}$, $C_{12}$, $C_{13}$, $C_{33}$, and $C_{44}$) of XB$_2$ (X = Cr, Mo and W) phases are summarized in Table 2. Mechanically stable phases should satisfy the well-known Born criteria [15,42,43] applicable to hexagonal structure:

$$C_{11} > 0, C_{44} > 0, (C_{11} - C_{12}) > 0, (C_{11} + C_{12})C_{33} - 2C_{12}^2 > 0 \qquad (9)$$

All the elastic constants of XB$_2$ (X = Cr, Mo and W) are positive and satisfy the above-mentioned criteria. This implies that XB$_2$ (X = Cr, Mo and W) compounds are mechanically stable.

The elastic tensors $C_{11}$ and $C_{33}$ signify the elastic stiffness of crystals against the strains along the principal axes (the crystallographic a- and c-directions, respectively). The difference between $C_{11}$ and $C_{33}$ implies that crystals are elastically anisotropic. In this view, the elastic anisotropy level in CrB$_2$ is more than that of MoB$_2$ and WB$_2$. The low value of $C_{44}$ indicates the shearability of the compounds. Of the three compounds studied here, WB$_2$ is the most resistant to shear deformation while the CrB$_2$ is the most shearable one. Additionally, the indentation hardness of materials is associated with the elastic tensor $C_{44}$. The indentation hardness increases with $C_{44}$. Thus, of the three phases discussed here, WB$_2$ is expected to be the hardest.

Table 2: Calculated elastic constants $C_{ij}$ of XB$_2$ (X = Cr, Mo and W) compounds (all in GPa).

| Compound | $C_{11}$ | $C_{12}$ | $C_{13}$ | $C_{33}$ | $C_{44}$ | Reference |
|---|---|---|---|---|---|---|
| CrB$_2$ | 559.90 | 120.79 | 166.03 | 309.62 | 126.20 | This work |
| | 583.70 | 117.30 | 119.30 | 343.30 | 143.10 | [24] |
| MoB$_2$ | 645.33 | 138.52 | 242.09 | 453.44 | 177.62 | This work |
| | 613.00 | 120.00 | 220.00 | 391.00 | 168.00 | [12] |
| | 621.40 | 116.90 | 228.40 | 404.30 | 175.00 | [15] |
| WB$_2$ | 717.28 | 256.20 | 330.88 | 668.87 | 202.14 | This work |
| | 590.10 | 187.40 | 236.40 | 442.80 | 98.80 | [23] |

The Hill approximated values of bulk modulus ($B_H$) and shear modulus ($G_H$) (using the Voigt-Reuss-Hill (VRH) method) [44–47], Young's modulus ($Y$), Poisson's ratio ($\sigma$), Lamé constants ($\mu$, $\lambda$), Tetragonal shear modulus ($C'$), Cauchy pressure, ($C''$), machinability index ($\mu_M$), Kleinman parameter ($\xi$) of XB$_2$ (X = Cr, Mo and W) have been computed using the following standard formulae [15,46,48–54]:

$$B_H = \frac{B_V + B_R}{2} \qquad (10)$$

$$G_H = \frac{G_V + G_R}{2} \qquad (11)$$

$$Y = \frac{9BG}{(3B + G)} \qquad (12)$$

$$\sigma = \frac{(3B - 2G)}{2(3B + G)} \qquad (13)$$

$$\mu = \frac{Y}{2(1 + \sigma)} \qquad (14)$$

$$\lambda = \frac{\sigma Y}{(1 + \sigma)(1 - 2\sigma)} \qquad (15)$$

$$C' = \frac{C_{11} - C_{12}}{2} \qquad (16)$$

$$C'' = (C_{12} - C_{44}) \qquad (17)$$

$$\mu_M = \frac{B}{C_{44}} \qquad (18)$$

$$\zeta = \frac{C_{11} + 8C_{12}}{7C_{11} + C_{12}} \qquad (19)$$

The results obtained are presented in Table 3 and Table 4. Table 3 shows that $B > G$, which implies that the shearing stress should determine the mechanical stability of $XB_2$ (X = Cr, Mo and W) compounds and the shape deforming stress, as opposed to the volume altering stress, will dictate the mechanical failure mechanism of $XB_2$ (X = Cr, Mo, and W) compounds. The value of $Y$ represents a material's stiffness (resistance to length change) [55,56]. Higher Young's modulus corresponds to increased stiffness of a material. Moreover, greater values of Young's modulus indicate a higher degree of covalency in a material [57]. The tetragonal shear modulus of a crystal is the measure of crystal's stiffness and its positive value suggests the dynamical stability of the solid. Therefore, it is anticipated that the compounds being studied will be dynamically stable.

Table 3: Bulk modulus B and shear modulus G (both in Voigt-Reuss-Hill (VRH) method), Young's modulus Y, tetragonal shear modulus C' and Cauchy pressure C'' of $XB_2$ (X = Cr, Mo and W) compounds (all in GPa).

| Compound | $B_V$ | $B_R$ | B | $G_V$ | $G_R$ | G | Y | C' | C'' | Reference |
|---|---|---|---|---|---|---|---|---|---|---|
| $CrB_2$ | 259.46 | 244.76 | 252.11 | 159.50 | 143.02 | 151.26 | 378.15 | 219.56 | -5.41 | This work |
| | - | - | 240.00 | - | - | 175.00 | 422.50 | - | - | [23] |
| MoB$_2$ | 332.17 | 329.77 | 330.97 | 196.49 | 181.69 | 189.09 | 476.52 | 253.41 | -39.10 | This work |
| | - | - | 304.00 | - | - | 186.00 | 463.00 | - | - | [12] |
| WB$_2$ | 437.71 | 437.55 | 437.63 | 205.99 | 202.87 | 204.43 | 530.67 | 230.54 | 54.06 | This work |
| | 327.00 | 324.10 | 325.60 | 144.00 | 129.50 | 136.70 | 359.70 | - | - | [22] |

Some of the indicators that are used to determine whether a solid is brittle or ductile include the Pugh's ratio (*B/G*), Poisson's ratio (*σ*), and Cauchy pressure (*C″*). The critical value of Pugh's ratio is 1.75 [58,59], Poisson's ratio is 0.26 [60], and Cauchy pressure is zero [51,53]. If the values of C″ are positive (negative), σ is higher (lower) than 0.26, and *B/G* is larger (lesser) than 1.75, the material is considered ductile (brittle) [38,61]. The calculated values of these parameters, which are shown in Table 3 and Table 4, suggest that CrB$_2$ and MoB$_2$ are brittle whereas WB$_2$ is ductile.

The Poisson ratio (*σ*) defines the relationship between lateral and axial strain caused by an applied axial stress [17,62]. For solids with central force dominance, the value of *σ* ranges from 0.25 to 0.50 [63]. For XB$_2$, the values of *σ* vary from 0.25–0.30, suggesting that the interatomic force is the central in type. In simple terms, the value of *σ* for covalent materials is around 0.10 and G ~1.1B; for ionic materials, it is usually about 0.25 and G ~0.6B; and for metallic materials, it is usually about 0.25 or more and G ~0.4B [64]. In our case the values of *σ* for XB$_2$ vary from about 0.25–0.30, indicating the delocalized metallic bonding should be dominant for the diborides. We have also calculated the Lam*é* constants. In terms of physical properties, the first Lam*é* constant, *λ* indicates the compressibility of the material while the second Lam*é* constant, *μ* reflects its shear stiffness [15].

The property of a material that determines how readily a cutting tool may be used to machine it is referred to as its "machinability". In today's industry, knowing a solid's machinability is crucial since it determines the best possible machine usage, cutting forces, temperature, and plastic strain. This indicator can also be used to test a solid's plasticity and dry lubricating properties [65–67]. A high value of machinability index ($\mu_M$) suggests greater ease of shape manipulation, lower feed forces, excellent dry lubricating properties, lower friction value, and higher plastic strain value. The values of machinability index of XB$_2$ indicate that they possess high level of machinability. The compounds also have a notable amount of dry lubricity.

The Kleinman parameter (*ζ*), commonly known as the internal strain parameter, measures a substance's ability to bend and stretch. The value of *ζ* typically varies from 0 to 1 [49]. The lower and higher limits of *ζ* signify the noteworthy role of bond stretching/contracting and bond bending to resist external force, respectively. The calculated values of *ζ* of XB$_2$ indicate that the bond stretching contribution dominates in WB$_2$, MoB$_2$ and CrB$_2$.

Table 4: Pugh's indicator or ratio $B/G$, Kleinman parameter $\zeta$, Poisson's ratio $\sigma$, Lamé constants ($\mu$, $\lambda$), and machinability index $\mu_M$ of $XB_2$ (X = Cr, Mo and W) compounds.

| Compound | $B/G$ | $\zeta$ | $\sigma$ | $\mu$ | $\lambda$ | $\mu_M$ | Reference |
|---|---|---|---|---|---|---|---|
| $CrB_2$ | 1.67 | 0.37 | 0.25 | 151.26 | 151.26 | 2.00 | This work |
|  | - | - | 0.20 | - | 199.60 | - | [15,23] |
| $MoB_2$ | 1.75 | 0.38 | 0.26 | 189.09 | 204.85 | 1.86 | This work |
|  | - | - | 0.28 | - | 241.80 | - | [15] |
| $WB_2$ | 2.14 | 0.50 | 0.30 | 204.10 | 306.16 | 2.16 | This work |
|  | - | - | 0.32 | - | 324.30 | - | [15,22] |

Table 4 presents the findings from earlier theoretical research [15,22,23]. Very good agreement with the calculated parameters is found.

Hardness is defined as the resistance of a material to local deformation induced by pressing a harder solid (indenter). From the standpoint of the application, the material's hardness is crucial. $C_{44}$ and $G$ are considered to be the best elastic constant and modulus for predicting the hardness of solids [68]. Hardness can be characterized theoretically by various schemes. X. Chen *et al.* ($H_{macro}$) [69], Y. Tian *et al.* [$(H_v)_{Tian}$] [70], and D. M. Teter [$(H_v)_{Teter}$] [71] developed hardness formulae based on either $G$ or both $G$ and $B$, whereas the formulae developed by N. Miao *et al.* ($H_{micro}$) [72] and E. Mazhnik *et al.* [$(H_v)_{Mazhnik}$] [73] are based on the Young's modulus and Poisson's ratio. The values of $H_{micro}$, $H_{macro}$, $(H_v)_{Tian}$, $(H_v)_{Teter}$, and $(H_v)_{Mazhnik}$ have been calculated using the following equations:

$$H_{micro} = \frac{(1-2\sigma)Y}{6(1+\sigma)} \qquad (20)$$

$$H_{macro} = 2[\left(\frac{G}{B}\right)^2 G]^{0.585} - 3 \qquad (21)$$

$$(H_V)_{Tian} = 0.92(G/B)^{1.137} G^{0.708} \qquad (22)$$

$$(H_V)_{Teter} = 0.151G \qquad (23)$$

$$(H_V)_{Mazhnik} = \gamma_0 \chi(\sigma) Y \qquad (24)$$

In Equation (24), $\chi(\sigma)$ is a function of the Poisson's ratio and can be evaluated from:

$$\chi(\sigma) = \frac{1 - 8.5\sigma + 19.5\sigma^2}{1 - 7.5\sigma + 12.2\sigma^2 + 19.6\sigma^3}$$

where $\gamma_0$ is a dimensionless constant with a value of 0.096. The discrepancy in hardness values obtained is due to the varied parameters used in equations (20) - (24). The obtained values are disclosed in Table 5. It appears that all the studied compounds possess high level of hardness.

In heavy-duty equipment, the formation of cracks is one of the most significant issues with the surface hard coatings, particularly in ceramic and metal materials. A solid's resistance to the initiation of cracks or fractures is measured by its fracture toughness, or $K_{IC}$. $K_{IC}$ has been evaluated using the following formula [74]:

$$K_{IC} = \alpha_0^{-1/2} V_0^{1/6} [\xi(\sigma)Y]^{3/2} \qquad (25)$$

where $V_0$ = volume per atom; $\alpha_0$ = 8840 GPa (for covalent and ionic crystals); $\xi(\sigma)$ is a parameter depending on the Poisson's ratio ($\sigma$), which can be found from:

$$\xi(\sigma) = \frac{1 - 13.7\sigma + 48.6\sigma^2}{1 - 15.2\sigma + 70.2\sigma^2 - 81.5\sigma^3}$$

As presented in Table 5, the values of $K_{IC}$ of the $XB_2$ (X = Cr, Mo and W) are 3.28, 5.38, and 5.13 MPam$^{1/2}$, respectively. It is evident that $MoB_2$ provides the best resistance against the formation and propagation of surface cracks. This result agrees with the computed hardness value.

Table 5: Calculated hardness (GPa) based on elastic moduli and Poisson's ratio and fracture toughness $K_{IC}$ (MPam$^{1/2}$) of $XB_2$ (X = Cr, Mo and W) compounds.

| Compound | $(H_V)_{Micro}$ | $(H_V)_{macro}$ | $(H_V)_{Tian}$ | $(H_V)_{Teter}$ | $(H_V)_{Mazhnik}$ | $K_{IC}$ | Reference |
|---|---|---|---|---|---|---|---|
| $CrB_2$ | 25.21 | 17.73 | 17.98 | 22.84 | 17.43 | 3.28 | This work |
| $MoB_2$ | 30.26 | 19.31 | 19.92 | 28.55 | 22.58 | 5.38 | This work |
| | - | 15.90 | - | - | - | - | [12] |
| $WB_2$ | 27.54 | 15.45 | 16.74 | 30.87 | 27.67 | 5.13 | This work |
| | 26.50 | - | - | - | - | - | [75] |

The elastic anisotropy of crystal refers to a difference in the elastic responses upon loading along different directions. Additionally, elastic anisotropy influences the way in which microscale cracks in ceramics grow and propagate, as well as the mobility of charged defects, the alignment or misalignment of quantum dots, unusual phonon modes, plastic relaxation in thin films, and other elastic processes. The anisotropy levels of $XB_2$ (X = Cr, Mo and W) are computed using the widely employed formulas.

For a hexagonal crystal, the shear anisotropy factor may be expressed in terms of three distinct parameters [76,77]. Taking into account the {100} shear planes between the ⟨011⟩ and ⟨010⟩ directions, the shear anisotropy factor, $A_1$ is as follows:

$$A_1 = \frac{4C_{44}}{C_{11}+C_{33}-2C_{13}} \qquad (26)$$

Taking into account the {010} shear plane between ⟨101⟩ and ⟨001⟩ directions the shear anisotropy factor, $A_2$ is as follows:

$$A_2 = \frac{4C_{55}}{C_{22}+C_{33}-2C_{23}} \qquad (27)$$

Taking into account the {001} shear planes between ⟨110⟩ and ⟨010⟩ directions, the anisotropy factor, $A_3$ is as follows:

$$A_3 = \frac{4C_{66}}{C_{11}+C_{22}-2C_{12}} \qquad (28)$$

The deviations of the factors $A_1$, $A_2$, and $A_3$ from unity reflect the crystal's elastic anisotropy for shape-changing deformation. Table 6 shows the calculated values for these anisotropy factors. The $XB_2$ (X = Cr, Mo, and W) compounds are predicted to be anisotropic based on the calculated values of $A_1$, $A_2$, and $A_3$.

The elastic coefficients provide a method to evaluate the linear compressibility. The formula for defining the compressibility anisotropy coefficients $k_c/k_a$, which are specific to hexagonal crystals exclusively, is the ratio of the linear compressibility coefficient along the crystallographic c-axis to the linear compressibility coefficient along the crystallographic a-axis and can be expressed as [78]:

$$k_c/k_a = \frac{(C_{11}+C_{12}-2C_{13})}{(C_{33}-C_{13})} \qquad (29)$$

Deviation of $k_c/k_a$ from unity determines the anisotropy level in linear compression. The $k_c/k_a$ of $CrB_2$, $MoB_2$, and $WB_2$ is about 2.43, 1.42 and 0.92, respectively. These values suggested that the compressibility along the c-axis is much greater than the a-axis in $CrB_2$ and $MoB_2$ compounds while the compressibility along the c-axis is much smaller than the a-axis in $WB_2$ which agrees well with the calculated elastic constants along different axis shown in Table 2.

The universal log-Euclidean anisotropy index is defined as follows [79,80]:

$$A^L = \sqrt{[\ln(\frac{B_V}{B_R})]^2 + 5[\ln(\frac{C_{44}^V}{C_{44}^R})]^2} \qquad (30)$$

In this scheme, the Voigt and Reuss values of $C_{44}$ are calculated from [79]:

$$C_{44}^R = \frac{5}{3} \frac{C_{44}(C_{11} - C_{12})}{3(C_{11} - C_{12}) + 4C_{44}}$$

and

$$C_{44}^V = C_{44}^R + \frac{3}{5} \frac{(C_{11} - C_{12} - 2C_{44})^2}{3(C_{11} - C_{12}) + 4C_{44}}$$

For perfect isotropy, $A^L = 0$. It is challenging to determine if a solid is layered or lamellar based solely on the value of $A^L$. Generally, large number (78%) of inorganic crystalline solids that have high $A^L$ values, exhibit layered/lamellar structure [79]. Roughly, compounds with higher (lower) $A^L$ values exhibit significant layered (non-layered) structural character. For the low values of $A^L$, $XB_2$ compounds are likely to have non-layered features.

The universal anisotropy index, $(A^U, d_E)$, equivalent Zener anisotropy measure, $A^{eq}$, percentage anisotropy in compressibility, $A_B$ and anisotropy in shear, $A_G$ (or $A_C$) for crystals are calculated from the following expressions [81–83]:

$$A^U = 5 \frac{G_V}{G_R} + \frac{B_V}{B_R} - 6 \geq 0 \qquad (31)$$

$$d_E = \sqrt{A^U + 6} \qquad (32)$$

$$A^{eq} = \left(1 + \frac{5}{12}A^U\right) + \sqrt{(1 + \frac{5}{12}A^U)^2 - 1} \qquad (33)$$

$$A_B = \frac{B_V - B_R}{B_V + B_R} \qquad (34)$$

$$A_G = \frac{G_V - G_R}{2G_H} \qquad (35)$$

A unique way to quantify anisotropy that works for all crystal systems regardless of their symmetry is the universal anisotropy index, $A^U$. Isotropy is implied by $A^U = 0$, whereas anisotropy is shown by a nonzero value of $A^U$. Anisotropy is predicted for $XB_2$ (X = Cr, Mo and W) compounds based on the calculated values of $A^U$. On the other hand, equivalent Zener anisotropy measure, $A^{eq} = 1$, represents isotropy, while any other value suggests anisotropy. Table 6 displays the computed values of $A^{eq}$ for $XB_2$, which predicts that $XB_2$ are elastically anisotropic. Two anisotropy variables ($A_B$ and $A_G$, respectively) that are proportionate to anisotropy in compression and shear exist. The factors are called percentage anisotropy factors, and they provide zero values for crystals that are perfectly isotropic when taking shear and compression into account. The degree of anisotropy increases when the value increases from zero to a positive value. For $XB_2$, anisotropy in shear is stronger than anisotropy in compressibility, as indicated by the larger value of $A_G$ (Table 6) in comparison to $A_B$. Overall, it appears that $CrB_2$ is more anisotropic than $MoB_2$ and $WB_2$. $WB_2$ is least anisotropic among the three.

Table 6: Shear anisotropy factors ($A_1$, $A_2$ and $A_3$), linear compressibility coefficients $k_c/k_a$, universal log-Euclidean index $A^L$, universal anisotropy index $A^U$, equivalent Zener anisotropy measure $A^{eq}$, anisotropy in compressibility $A^B$ (%), and anisotropy in shear $A^G$ or $A^C$ (%) of $XB_2$ (X = Cr, Mo and W) compounds.

| Compound | $A_1$ | $A_2$ | $A_3$ | $k_c/k_a$ | $A^L$ | $A^U$ | $A^{eq}$ | $A^B$ | $A^G$ | Layered/Non-layered |
|---|---|---|---|---|---|---|---|---|---|---|
| $CrB_2$ | 0.94 | 0.94 | 1.00 | 2.43 | 0.46 | 0.64 | 2.04 | 0.03 | 0.05 | Non-layered |
| $MoB_2$ | 1.16 | 1.16 | 1.00 | 1.42 | 0.20 | 0.41 | 1.78 | 0.004 | 0.04 | Non-layered |
| $WB_2$ | 1.12 | 1.12 | 1.00 | 0.92 | 0.03 | 0.08 | 1.29 | 0.002 | 0.008 | Non-layered |

We have analyzed and displayed the two-dimensional (2D) and three-dimensional (3D) profiles of the Young's modulus ($Y$), shear modulus ($G$), compressibility ($\beta$) (inverse of the bulk modulus), and Poisson's ratio ($\sigma$) of the $XB_2$ (X = Cr, Mo and W) compounds. These profiles were generated by ELATE [84]. Mechanically isotropic solids require 3D contour plots with spherical forms; otherwise, anisotropy exists. Figure 2 displays the 3D profiles of $Y$, $\beta$, $G$, and $\sigma$, all of which deviate from the spherical form, demonstrating anisotropy. ELATE also offers a numerical study, displaying the minimum and maximum values of each modulus, as well as the directions along which these extrema occur. The maximum and minimum values of $Y$, $\beta$, $G$, and $\sigma$, as well as their respective maximum to minimum ratios are disclosed in Table 7. These ratios are important in determining the elastic anisotropy in $XB_2$. The anisotropy $A_X$ of each elastic modulus $X$ can be defined as follows [43,84]:

$$A_X = \begin{Bmatrix} X_{max}/X_{min} & if\ sign(X_{max}) = sign(X_{min}) \\ \infty & otherwise. \end{Bmatrix} \qquad (36)$$

The anisotropy in Young's modulus, linear compressibility, shear modulus and Poisson's ratio are observed to be the lowest for $WB_2$ and highest for $CrB_2$. The 2D plots depict the projections of 3D profiles onto the principal crystal planes. These plots show how the elastic properties exhibit anisotropy in various directions inside the specified crystal planes.

The plots for shear modulus and Poisson's ratio are complex due to their dependence on two orthogonal unit vectors in the direction of the measurement and the direction of applied stress [84,85]. Figures 4 and 5 show the 2D projections of shear modulus and Poisson's ratio, with green (blue) lines representing the minimum (maximum) positive values of each property of the respective elastic parameters.

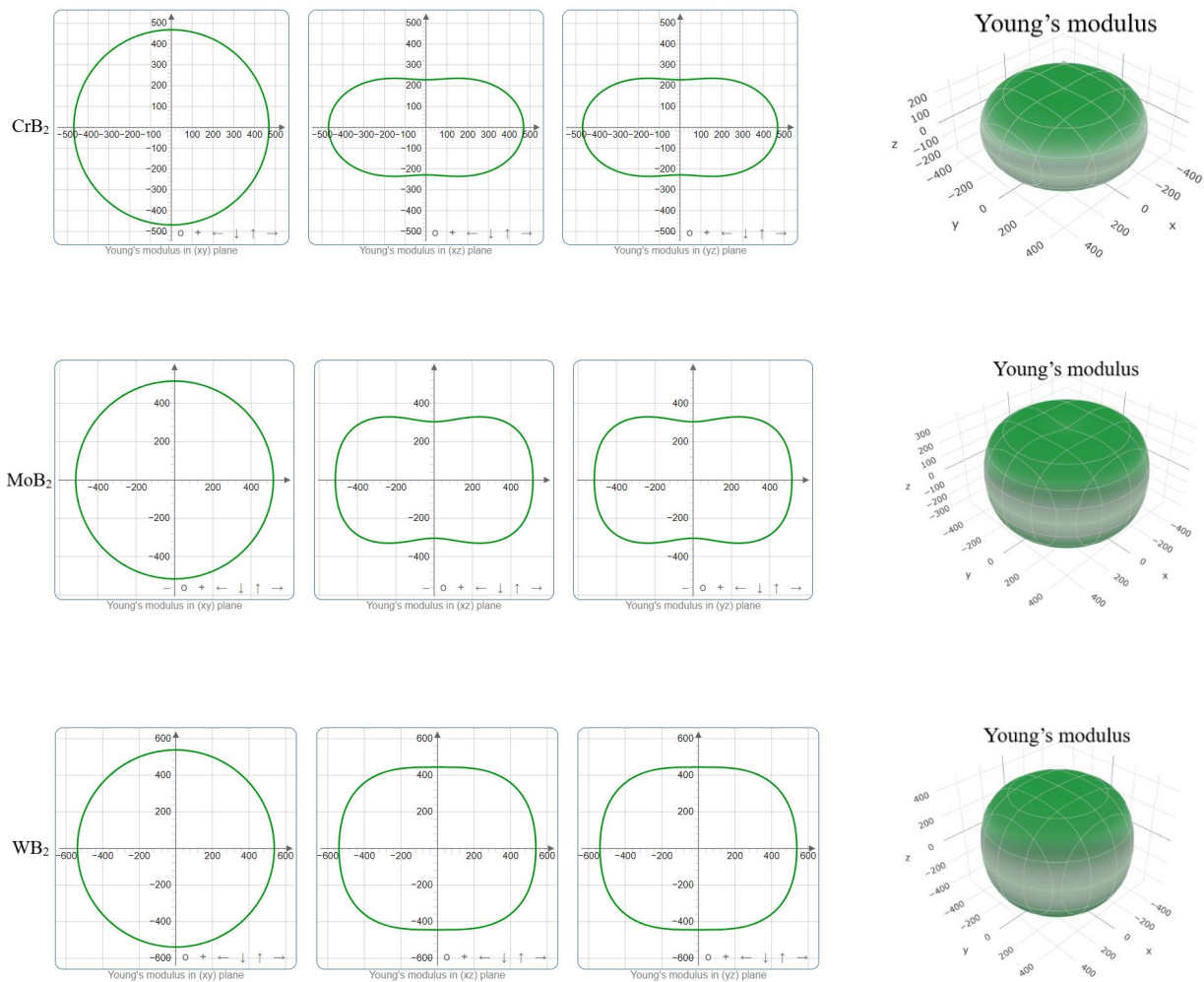

**Figure 2.** 2D (left) and 3D (right) directional dependences in Young's modulus ($Y$) of XB$_2$ compounds.

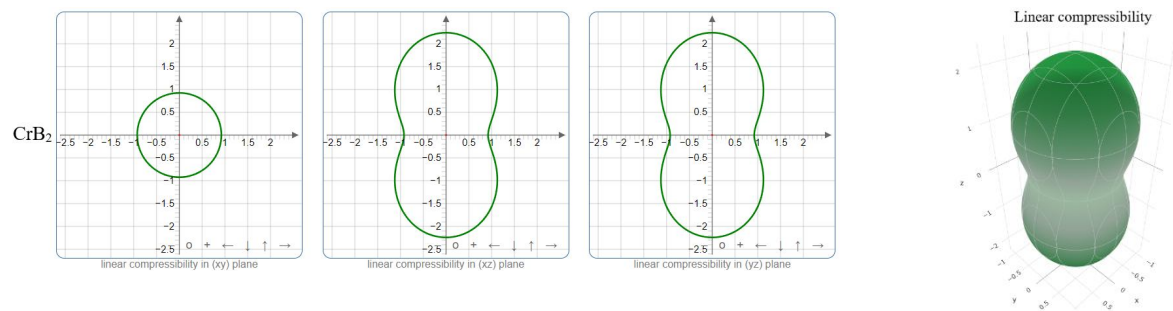

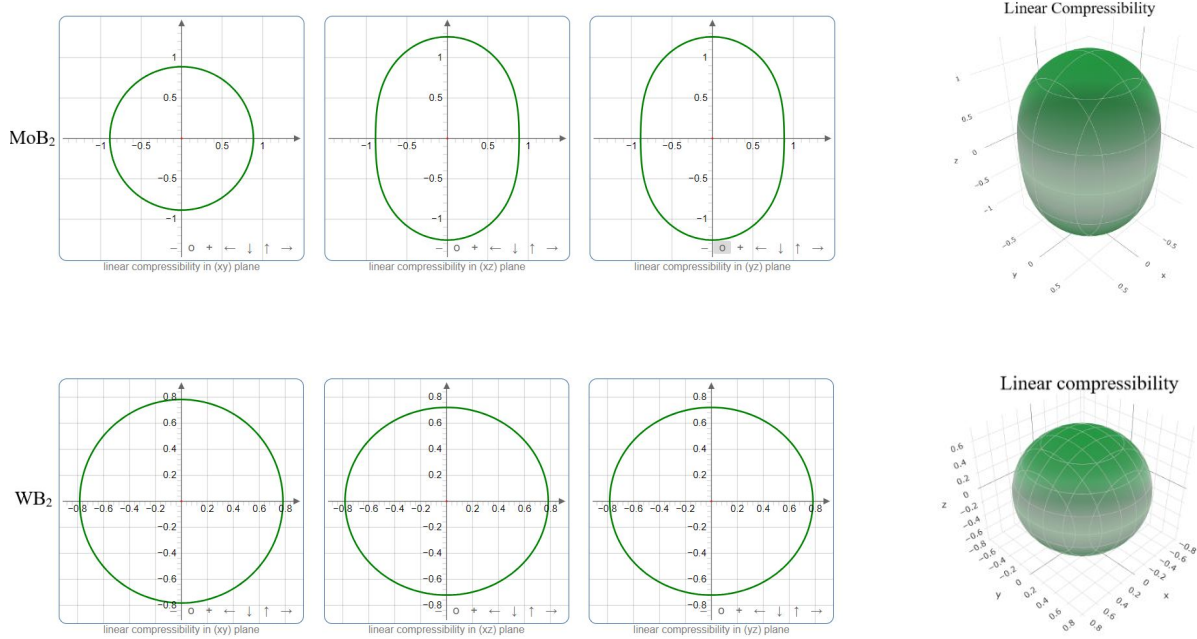

**Figure 3.** 2D (left) and 3D (right) directional dependences in compressibility (*β*) of XB$_2$ compounds.

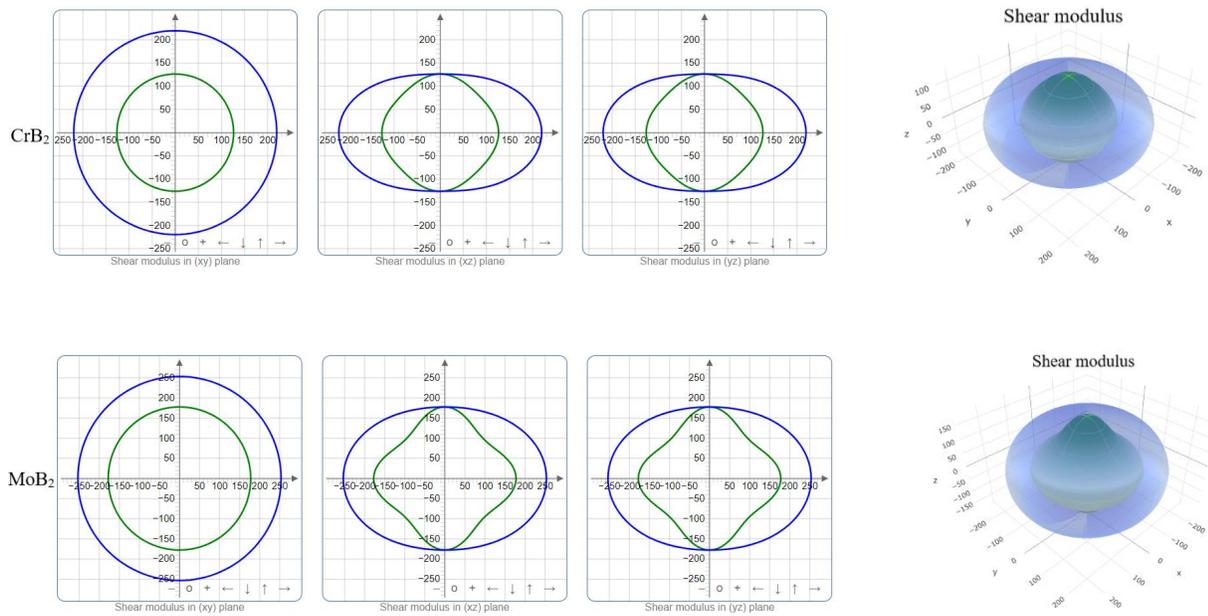

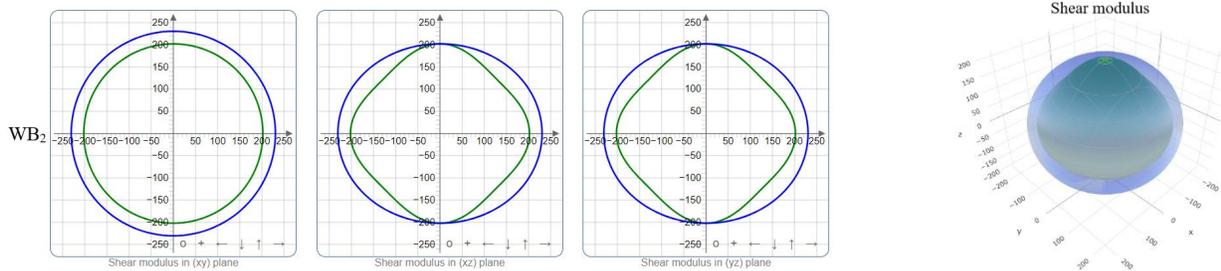

**Figure 4.** 2D (left) and 3D (right) directional dependences in shear modulus (*G*) of XB$_2$ compounds.

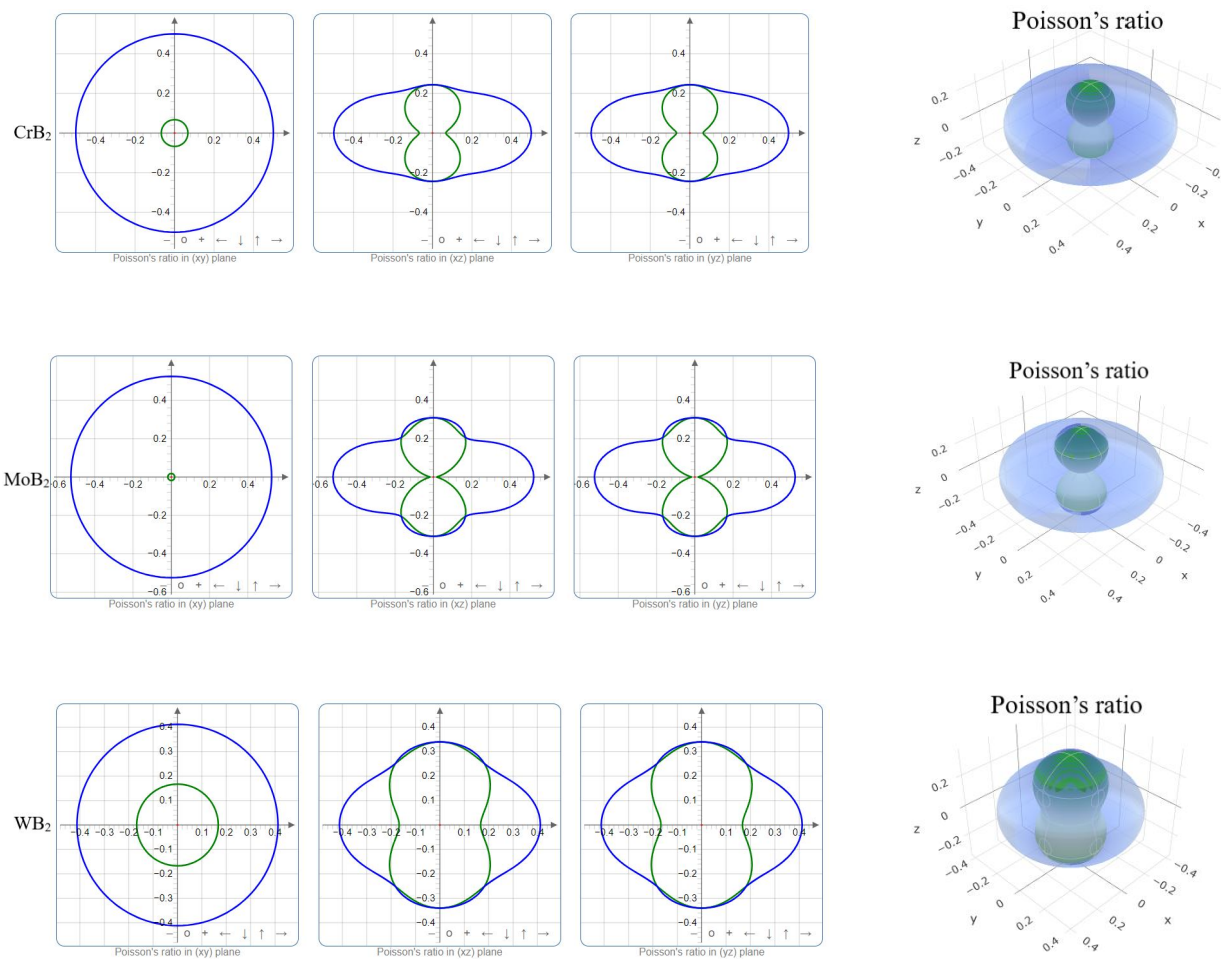

**Figure 5.** 2D (left) and 3D (right) directional dependences in Poisson's ratio ($\sigma$) of XB$_2$ compounds.

Table 7: The minimum and maximum values of Young's modulus (GPa), compressibility (TPa$^{-1}$), shear modulus (GPa), Poisson's ratio, and elastic anisotropies $A_X$ ($X = Y, \beta, G, \sigma$) of XB$_2$ compounds.

| Phase | Y | | $A_Y$ | $\beta$ | | $A_\beta$ | G | | $A_G$ | $\sigma$ | | $A_\sigma$ |
|---|---|---|---|---|---|---|---|---|---|---|---|---|
| | $Y_{min}$ | $Y_{max}$ | | $\beta_{min}$ | $\beta_{max}$ | | $G_{min}$ | $G_{max}$ | | $\sigma_{min}$ | $\sigma_{max}$ | |
| CrB$_2$ | 228.63 | 468.73 | 2.05 | 0.92 | 2.24 | 2.43 | 115.73 | 219.56 | 1.89 | 0.07 | 0.50 | 7.41 |
| MoB$_2$ | 303.90 | 519.11 | 1.71 | 0.89 | 1.26 | 1.42 | 137.71 | 253.41 | 1.84 | 0.02 | 0.52 | 29.21 |
| WB$_2$ | 443.95 | 547.72 | 1.23 | 0.72 | 0.78 | 1.08 | 177.24 | 230.54 | 1.30 | 0.17 | 0.41 | 2.47 |

### 3.3 Electronic band structure and density of states

To understand the electronic and optical properties of XB$_2$ (X = Cr, Mo and W) compounds, an investigation of the electronic band structure is essential. The behavior of electrons within the Brillouin zone (BZ) is determined by their energy dispersion [$E(k)$]. We have calculated the electronic band structure for the optimized crystal structures of XB$_2$ (X = Cr, Mo and W) compounds along several high symmetry directions ($\Gamma$-M-K-$\Gamma$-A-L-H-A) in the first BZ. The Fermi level is indicated by the horizontal broken line placed at zero energy. As shown in Figure 6, no band gap exists in the band structures exhibiting the metallic character. The presence of highly dispersed bands leads in a low charge carrier effective mass [86–88] and a high charge mobility. The scenario is reversed for the non-dispersive bands. Overall, for all the three metallic diborides under investigation, the bands along *L-H* directions are less dispersive indicating relatively high effective mass and as a result, low mobility of charge carriers along this direction. Overall, the degree of dispersion in the bands crossing the Fermi level is higher for WB$_2$ compared to the other two metallic diborides.

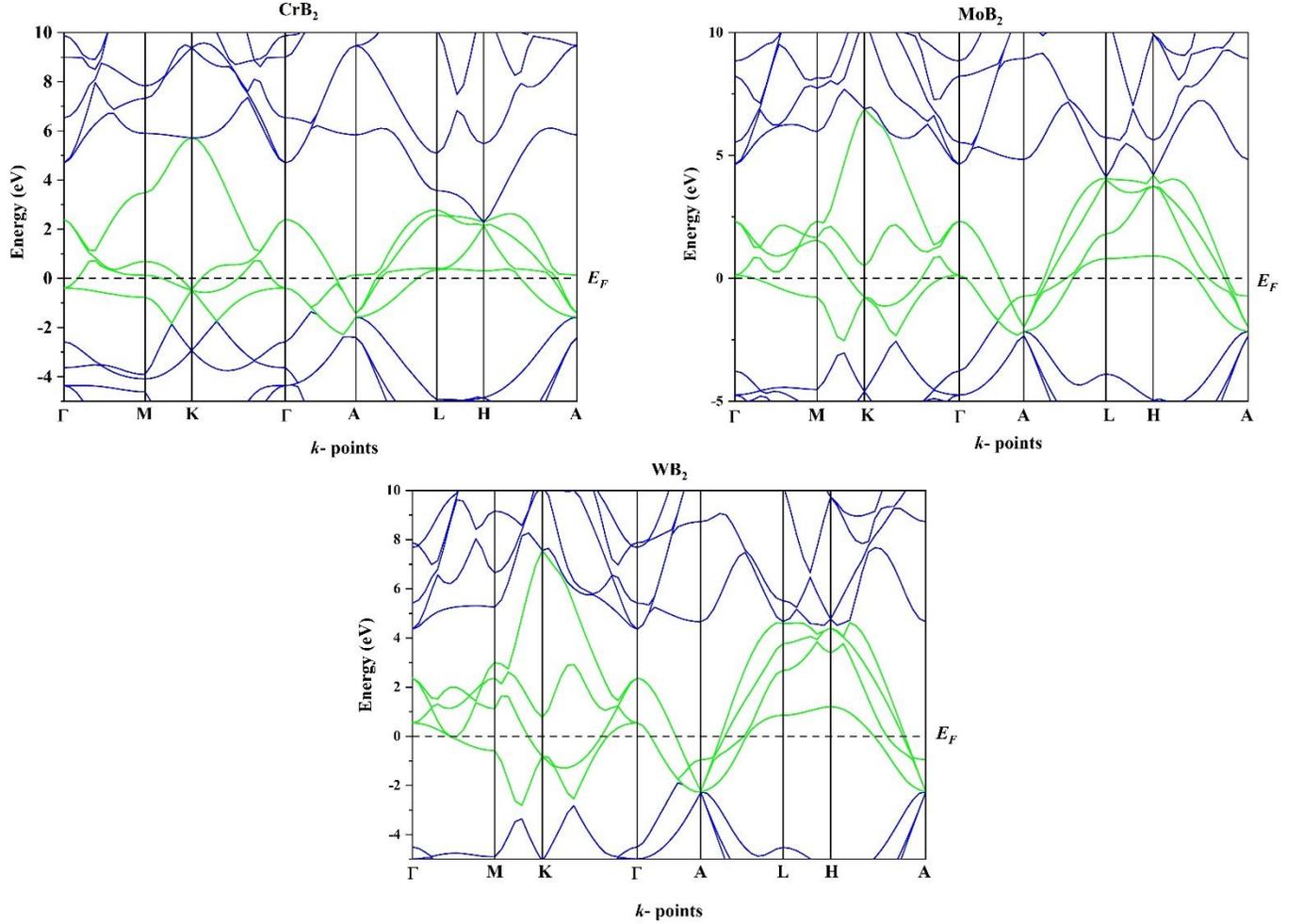

**Figure 6.** The band structures of $XB_2$ (X = Cr, Mo and W) compounds in the first Brillouin zone.

To understand the electronic properties of these three materials on a fundamental level, the total and partial density of states (TDOSs and PDOSs, respectively) are illustrated in Figure 7 in which vertical dashed line at 0 eV represents the Fermi level, $E_F$. To investigate the contribution of different atoms to the TDOS, we have computed the PDOSs for Cr, Mo, W, and B in $XB_2$ (X = Cr, Mo and W) compounds. At the Fermi level, the TDOSs for $CrB_2$, $MoB_2$, and $WB_2$ are 2.52, 1.34, and 1.25 electronic states per eV, respectively. As a result, of the three diborides, CrB2 should have the highest conductivity. Close to the Fermi level, the main contribution to the TDOSs comes from the Cr-$3d$ (84.92%) orbitals for $CrB_2$, Mo-$4d$ (74.18%) for $MoB_2$, and W-$5d$ (69.36%) for $WB_2$. Thus, these electronic states should dominate the electrical conductivity of the $XB_2$ compounds. The properties of this electronic state also have an impact on the chemical and mechanical stabilities of $XB_2$. The peaks, particularly those at the Fermi level, should influence charge flow and electrical transport properties. In the TDOS, the bonding peak is the nearest peak at the negative energy below the Fermi level, while the anti-bonding peak is the nearest peak at the positive energy above the Fermi energy. The energy gap between these peaks is called the pseudo-gap, which indicates electrical stability; crystals with more bonding electrons are structurally more stable [89,90]. In $XB_2$ compounds bonding and anti-bonding peaks are located within 2.70 eV for CrB$_2$, 0.84 eV for MoB$_2$, and 0.81eV for WB$_2$ from the Fermi level.

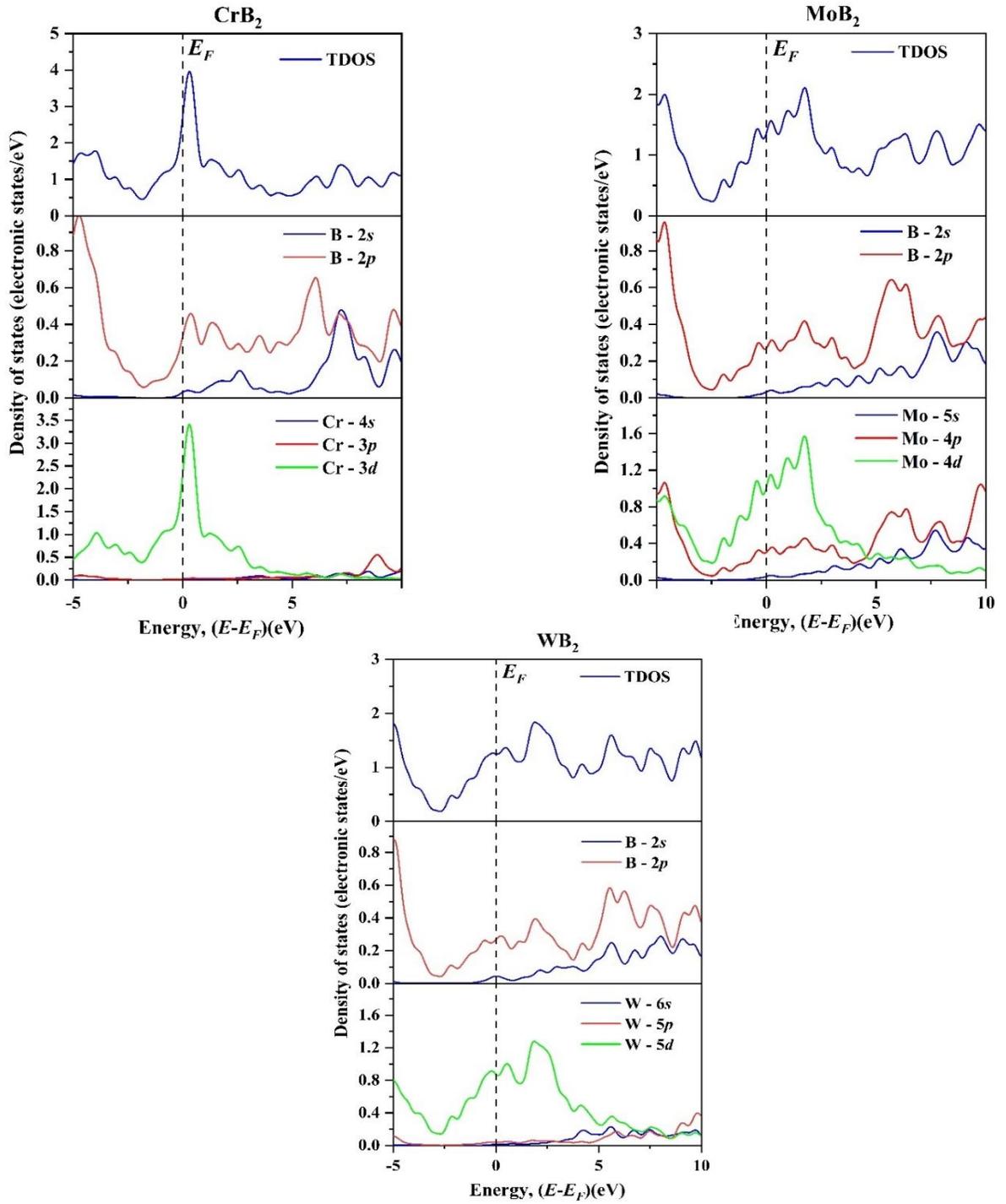

**Figure 7.** Total and partial electronic energy density of states of XB$_2$ (X = Cr, Mo and W) compounds.

### 3.4 Thermo-mechanical properties

(a) Acoustic behavior and its anisotropy

The relationship between sound velocity and thermal conductivity, $K = \frac{1}{3}C_v l v$ (where, $C_v$ is the specific heat per unit volume, $v$ is the velocity of sound in the solid, and $l$ is the mean free path for the lattice vibrations), makes sound velocity an important parameter of a material. In this section, the longitudinal and transverse sound velocities of $XB_2$ (X = Cr, Mo and W) have been calculated. Given the bulk ($B$) and shear ($G$) moduli, the longitudinal ($v_l$) and the transverse ($v_t$) sound velocities can be computed using the following equations [91]:

$$v_t = \sqrt{\frac{G}{\rho}} \qquad (37)$$

and

$$v_l = \sqrt{\frac{3B + 4G}{3\rho}} \qquad (38)$$

Here, $\rho$ is the mass-density of the solid. These formulae demonstrate that the density and elastic moduli of a material greatly affects its sound velocities. The average sound velocity, $v_a$, can be calculated using the longitudinal ($v_l$) and the transverse ($v_t$) sound velocities as follows [91]:

$$v_a = \left[\frac{1}{3}\left(\frac{2}{v_t^3} + \frac{1}{v_l^3}\right)\right]^{-\frac{1}{3}} \qquad (39)$$

The estimated values of sound velocities of $XB_2$ are represented in Table 8. The average sound velocity of $CrB_2$ is more than that of $MoB_2$ and $WB_2$. So, the compound, $CrB_2$ may behave as better room temperature heat conductor than $MoB_2$ and $WB_2$.

The study of acoustic impedance of materials has developed over time for many underwater acoustic applications, aircraft engine design, noise reduction, transducer design, and industrial factory design. A high sound pressure will produce a high particle velocity in a medium with low impedance, but the same sound pressure will produce a relatively modest particle velocity in a medium with high impedance. The acoustic impedance of the compounds was determined using the following equation [92]:

$$Z = \sqrt{\rho G} \qquad (40)$$

where $G$ is the shear modulus and $\rho$ is the density of the solid. The unit of acoustic impedance is the Rayl; 1 Rayl = $kgm^{-2}s^{-1}$ = 1 $Nsm^3$. An acoustic impedance of a material is high when its shear modulus and density are both high.

The intensity of sound radiation is another important factor in the design of sound boards and loudspeakers. The density and shear modulus of a material are related to the intensity, *I*, of its acoustic radiation as [92]:

$$I \approx \sqrt{G/\rho^3} \qquad (41)$$

This is also known as the *radiation factor*. In sound board design, the radiation factor is usually taken into consideration while selecting materials.

Table 8: Calculated mass density $\rho$ (kg/m$^3$), longitudinal velocity of sound $v_l$ (ms$^{-1}$), transverse velocity of sound $v_t$ (ms$^{-1}$), average sound velocity $v_a$ (ms$^{-1}$), acoustic impedance Z (Rayl) and radiation factor I (m$^4$/kg.s) of XB$_2$ (X = Cr, Mo and W) compounds.

| Compound | P | $v_l$ | $v_t$ | $v_a$ | Z ($\times 10^6$) | I |
|---|---|---|---|---|---|---|
| CrB$_2$ | 5437.49 | 9134.55 | 5273.76 | 5854.89 | 28.68 | 0.97 |
| MoB$_2$ | 7695.13 | 8704.82 | 4957.08 | 5509.82 | 38.15 | 0.64 |
| WB$_2$ | 1342.11 | 7273.67 | 3902.44 | 4357.78 | 52.39 | 0.29 |

Every atom in a solid can vibrate in one of three separate modes (one longitudinal and two transverse). In anisotropic crystals, pure longitudinal and transverse wave modes are limited only to certain crystallographic directions, whereas quasi-transverse or quasi-longitudinal modes exist in all other directions. Hexagonal symmetric crystals can exhibit pure longitudinal and transverse modes for [001] and [100] directions. The acoustic velocities of hexagonal crystals along these principal directions can be determined using the following relationships [83]:

$$[100]\upsilon_l = \sqrt{(C_{11} - C_{12})/2\rho}; \; [010]\upsilon_{t1} = \sqrt{C_{11}/\rho}; \; [001]\upsilon_{t2} = \sqrt{C_{44}/\rho} \qquad (42)$$

$$[001]\upsilon_l = \sqrt{C_{33}/\rho}; \; [100]\upsilon_{t1} = [010]\upsilon_{t2} = \sqrt{C_{44}/\rho} \qquad (43)$$

where $\upsilon_{t1}$ and $\upsilon_{t2}$ refers to the velocity of first transverse mode and the second transverse mode, respectively, and $\upsilon_l$ is the velocity of the longitudinal mode. These correlations predict that a compound with high elastic constants and low density will have high sound velocities. Directional sound velocities are displayed in Table 9.

Table 9: Anisotropic sound velocities (ms$^{-1}$) of XB$_2$ (X = Cr, Mo and W) compounds along different crystallographic directions.

| Propagation directions | | Anisotropic sound velocities (ms$^{-1}$) for compounds | | |
|---|---|---|---|---|
| | | CrB$_2$ | MoB$_2$ | WB$_2$ |
| [100] | [100]$v_l$ | 6353.8 | 5738.5 | 4144.1 |
| | [010]$v_{t1}$ | 10146.5 | 9157.6 | 7309.8 |
| | [001]$v_{t2}$ | 4817.1 | 4808.4 | 3880.5 |
| [001] | [001]$v_l$ | 7545.3 | 7676.3 | 7058.8 |
| | [100]$v_{t1}$ | 4817.1 | 4808.4 | 3880.5 |
| | [010]$v_{t2}$ | 4817.1 | 4808.4 | 3880.5 |

Significant anisotropy in sound velocity is present in XB$_2$ reflecting anisotropy in the bonding strengths along different crystallographic axes.

(b) Debye temperature

The Debye temperature ($\Theta_D$) is linked to various physical properties e.g., lattice vibration, thermal conductivity, phonon-specific heat, interatomic bonding, resistivity, melting temperature, vacancy formation energy, and coefficient of thermal expansion. In conventional superconductors, it also provides the characteristic energy of the phonons that leads to Cooper pairing. It is generally observed that materials with larger Debye temperatures are those with stronger interatomic bonding strength, higher melting temperature, higher hardness, higher acoustic wave velocity, and lower average atomic mass. Moreover, Debye temperature marks the boundary between the classical and quantum behavior in lattice vibration. When $T > \Theta_D$, all the modes of vibrations have almost the same energy, $\sim k_B T$. On the other hand, for $T < \Theta_D$, higher frequency modes are regarded frozen, revealing the quantum mechanical nature of vibrational energy spectrum [93]. In this study, $\Theta_D$ was computed as follows [91,94]:

$$\Theta_D = \frac{h}{k_B}\left[\left(\frac{3n}{4\pi}\right)\frac{N_A \rho}{M}\right]^{\frac{1}{3}} v_a \qquad (44)$$

where $h$ is the Planck's constant, $k_B$ is the Boltzmann's constant, $n$ is the number of atoms in the unit cell, $M$ is the molar mass, $\rho$ is the density of the solid, $N_A$ is the Avogadro number, and $v_a$ denotes the average sound velocity.

As shown in Table 10, $\varTheta_D$ value of $CrB_2$ is higher than that of $MoB_2$ and $WB_2$. The estimated values of $\varTheta_D$ of $XB_2$ (X = Cr, Mo and W) compounds exhibit excellent agreement with the earlier findings.

(c) Melting temperature

Melting point is a key thermo-physical factor useful for verifying the feasibility of using solids at high temperatures. A high melting temperature indicates strong atomic bonding, a high heat of fusion, low entropy of fusion, or a combination of both, as well as a low thermal expansion value [96]. We have calculated the melting point ($T_m$) of hexagonal crystals using the elastic stiffness constants from an empirical formula proposed by Fine et al. [95]:

$$T_m = 354\ K + 4.5(K/GPa)\left(\frac{2C_{11} + C_{33}}{3}\right) \qquad (45)$$

The calculated melting temperatures are enlisted in Table 10. All three $XB_2$ (X = Cr, Mo and W) compounds possess high melting point. Generally speaking, a solid has a greater melting temperature when its Young's modulus is higher, and vice versa [96,97]. Because of higher elastic constants ($C_{11}$ and $C_{33}$) and Young's modulus of $WB_2$, the melting point of $WB_2$ is higher than that of $MoB_2$ and $CrB_2$, which means $WB_2$ is a better candidate material for high temperature application compared to $MoB_2$ and $CrB_2$.

(d) Thermal expansion coefficient and heat capacity

The thermal expansion coefficient ($\alpha$) (TEC) is another important thermomechanical parameter controlled by the bonding strength and lattice anharmonicity. This particular parameter is linked to a variety of other physical properties, including specific thermal conductivity, heat, temperature variation of the energy band gap in semiconductors, and effective mass of electron/hole. The thermal expansion coefficient of a solid can be calculated using the following equation [93]:

$$\alpha = \frac{1.6 \times 10^{-3}}{G} \qquad (46)$$

The thermal expansion coefficient is inversely related to the melting temperature: $\alpha \approx 0.02/T_m$ [92,98]. The computed TECs are disclosed in Table 10.

The heat capacity per unit volume ($\rho C_P$) is the change in thermal energy per unit volume of a substance per degree Kelvin change in temperature. Materials with higher heat capacity exhibit lower thermal diffusivity and higher thermal conductivity. These materials have a high capacity for heat storage. We have calculated $\rho C_P$ for $XB_2$ (X = Cr, Mo and W) compounds using the following equation [92]:

$$\rho C_P = \frac{3k_B}{\Omega} \qquad (47)$$

where, $(1/\Omega)$ is the number of atoms per unit volume. Table 10 shows the heat capacity per unit volume of metallic diborides. The heat capacity of CrB$_2$ is higher than those of MoB$_2$ and WB$_2$. It follows that CrB$_2$ should have a relatively higher thermal conductivity.

Table 10: Number of atoms per unit volume n (atoms/m$^3$), elastic Debye temperature $\Theta_D$ (K), melting temperature T$_m$ (K), thermal expansion coefficient α (K$^{-1}$), and heat capacity per unit volume $\rho C_\rho$ (JK$^{-1}$m$^{-3}$) of XB$_2$ (X = Cr, Mo and W) compounds.

| Compound | n (10$^{28}$) | $\Theta_D$ | T$_m$ | α (10$^{-5}$) | $\rho C_\rho$ (10$^6$) | Reference |
|---|---|---|---|---|---|---|
| CrB$_2$ | 13.34 | 890.84 | 2498.13 | 1.06 | 5.52 | This work |
|  | - | 941.00 | - | - | - | [17] |
| MoB$_2$ | 11.82 | 805.91 | 2970.15 | 0.85 | 4.89 | This work |
|  | - | 783.00 | - | - | - | [17] |
| WB$_2$ | 11.80 | 636.44 | 3509.15 | 0.78 | 4.88 | This work |
|  | - | 624.20 | - | - | - | [22] |

(e) Lattice thermal conductivity

The ability of a material to transport heat through lattice vibrations is measured by its lattice thermal conductivity ($k_{ph}$) at various temperatures. The lattice thermal conductivity plays a crucial role in many technological applications as a design parameter. For example, low thermal conductivity is preferred for thermoelectric (TE) conversions and for thermal barrier coating (TBC) materials. On the other hand, high lattice thermal conductivity materials find extensive application in heat sinks [99]. Slack developed an empirical formula to theoretically estimate the $k_{ph}$ [100]:

$$k_{ph} = A(\gamma) \frac{M_{av} \Theta_D^3 \delta}{\gamma^2 n^{2/3} T} \qquad (48)$$

In the above equation, $M_{av}$ is the average atomic mass in kg/mol, $\delta$ is the cubic root of average atomic volume in meter (m), $\Theta_D$ is the Debye temperature in K, $T$ is the absolute temperature in K, $n$ is the number of atoms in the conventional unit cell, and $\gamma$ is the acoustic Grüneisen parameter that measures the degree of anharmonicity of the phonons. A material with a low Grüneisen parameter has low phonon anharmonicity, resulting in high thermal conductivity. The dimensionless quantity $A(\gamma)$ may be computed from the Poisson's ratio [100]:

$$\gamma = \frac{3(1+v)}{2(2-3v)} \qquad (49)$$

The factor $A(\gamma)$, due to Julian [101], is calculated from:

$$A(\gamma) = \frac{5.720 \times 10^7 \times 0.849}{2 \times (1 - 0.514/\gamma + 0.228/\gamma^2)} \qquad (50)$$

The estimated lattice thermal conductivity at room temperature (300 K) and the Grüneisen parameter are given in Table 11. It is found that the compound MoB$_2$ shows higher $k_{ph}$ than that of CrB$_2$ and WB$_2$.

(f) Minimum thermal conductivity and its anisotropy

The phonon thermal conductivity of a substance reaches its lower limit at higher temperature, which is known as the minimum thermal conductivity of that material. In this case, the interatomic distance is taken to be the mean free path of the phonons. The following formula for determining the minimum thermal conductivity, $k_{min}$, of materials at high temperatures was developed by Clarke [98] using the quasi-harmonic Debye model:
:

$$k_{min} = k_B \upsilon_a (V_{atomic})^{-2/3} \qquad (51)$$

where, $k_B$ is the Boltzmann constant, $\upsilon_a$ is the average sound velocity and $V_{atomic}$ represents the cell volume per atom of the compound.

The calculated values of minimum thermal conductivity of the XB$_2$ (X = Cr, Mo and W) compounds are also tabulated in Table 11. The $k_{min}$ of CrB$_2$ is higher than that of MoB$_2$ and WB$_2$. Higher minimum thermal conductivity is seen in compounds with higher sound velocity and $\Theta_D$.

An elastically anisotropic material has anisotropic minimum thermal conductivity. Anisotropic thermal conductivity materials gain extensive application in thermal barrier coatings, thermo-electrics, heat shields, and heat spreading in electronic and optical device technologies. Anisotropy in minimum thermal conductivity is influenced by the different sound velocities corresponding to different crystallographic directions. The following minimum thermal conductivities are calculated along various directions using the Cahill model [102]:

$$k_{min} = \frac{k_B}{2.48} n^{2/3} (\upsilon_l + \upsilon_{t1} + \upsilon_{t2}) \qquad (52)$$

where, $n = N/V$ = number of atoms per unit volume. The minimum thermal conductivities of XB$_2$ (X = Cr, Mo and W) compounds along the [100] and [001] directions are shown in Table

11. The minimum thermal conductivities along different crystallographic axes of $XB_2$ compounds appear to be higher than the isotropic minimum thermal conductivity.

Table 11: Grüneisen parameter γ, lattice thermal conductivity $k_{ph}$ (W/m-K), minimum thermal conductivities $k_{min}$ in different crystallographic directions (all in W/m-K), and minimum thermal conductivity $k_{min}$ in Cahill's and Clark's method (both in W/m-K) of $XB_2$ (X = Cr, Mo and W) compounds.

| Compound | γ | $k_{ph}$ | [100]$k_{min}$ | [001]$k_{min}$ | $k_{min}$ | |
| --- | --- | --- | --- | --- | --- | --- |
| | | | | | Cahill | Clark |
| $CrB_2$ | 1.50 | 77.39 | 3.10 | 2.50 | 2.80 | 2.11 |
| $MoB_2$ | 1.55 | 90.60 | 2.64 | 2.32 | 2.48 | 1.83 |
| $WB_2$ | 1.76 | 57.87 | 2.05 | 1.98 | 2.02 | 1.45 |

**3.5 Optical parameters**

Optical parameters such as absorption coefficient $α(ω)$, optical conductivity $σ(ω)$, dielectric function $ε(ω)$, loss function $L(ω)$, reflectivity $R(ω)$, and refractive index $N(ω)$ [where $ω = 2πf$ is the angular frequency of the incident electromagnetic wave (EMW)] are crucial for analyzing optical response of a material to incident light [103–105]. All the optical parameters are intimately related to the electronic band structure. In this section, we have provided the findings of the computed optical properties of $XB_2$ (X = Cr, Mo, and W) compounds for incident EMW energies up to 30 eV with electric field polarizations along the [100] and [001] directions, as depicted in Figures (8-10). Since the binary metallic diborides $XB_2$ have hexagonal symmetry, the directions [100] and [001] of incident photon refers to directions of related electric field at right angles and parallel to crystallographic c-axis, respectively.

The absorption coefficient, $α(ω)$, tells us how much light with a certain energy may enter a material before being absorbed and aids in determining a substance's ideal solar energy conversion efficiency. Low absorption coefficient materials merely absorb light weakly; if they are thin enough, they will turn transparent at certain wavelengths. It also indicates the electrical properties of a material, such as whether it is insulating, metallic, or semiconducting. The variation of the absorption coefficient $α(ω)$ as a function of photon energy is shown in Figures 8(a), 9(a), and 10(a). The absorption coefficient of all three metallic diborides begins from 0 eV of photon energy, confirming that they have metallic electronic band structures. $XB_2$ compounds have high absorption coefficient in the ultraviolet (UV) region, ranging from 3.5 to 25.0 eV in the spectral region. $CrB_2$ and $MoB_2$ have larger peak values for [100] polarization than for [001] polarization, whereas WB$_2$ has higher peak values for [001] polarization than for [100] polarization, demonstrating optical anisotropy in absorption properties.

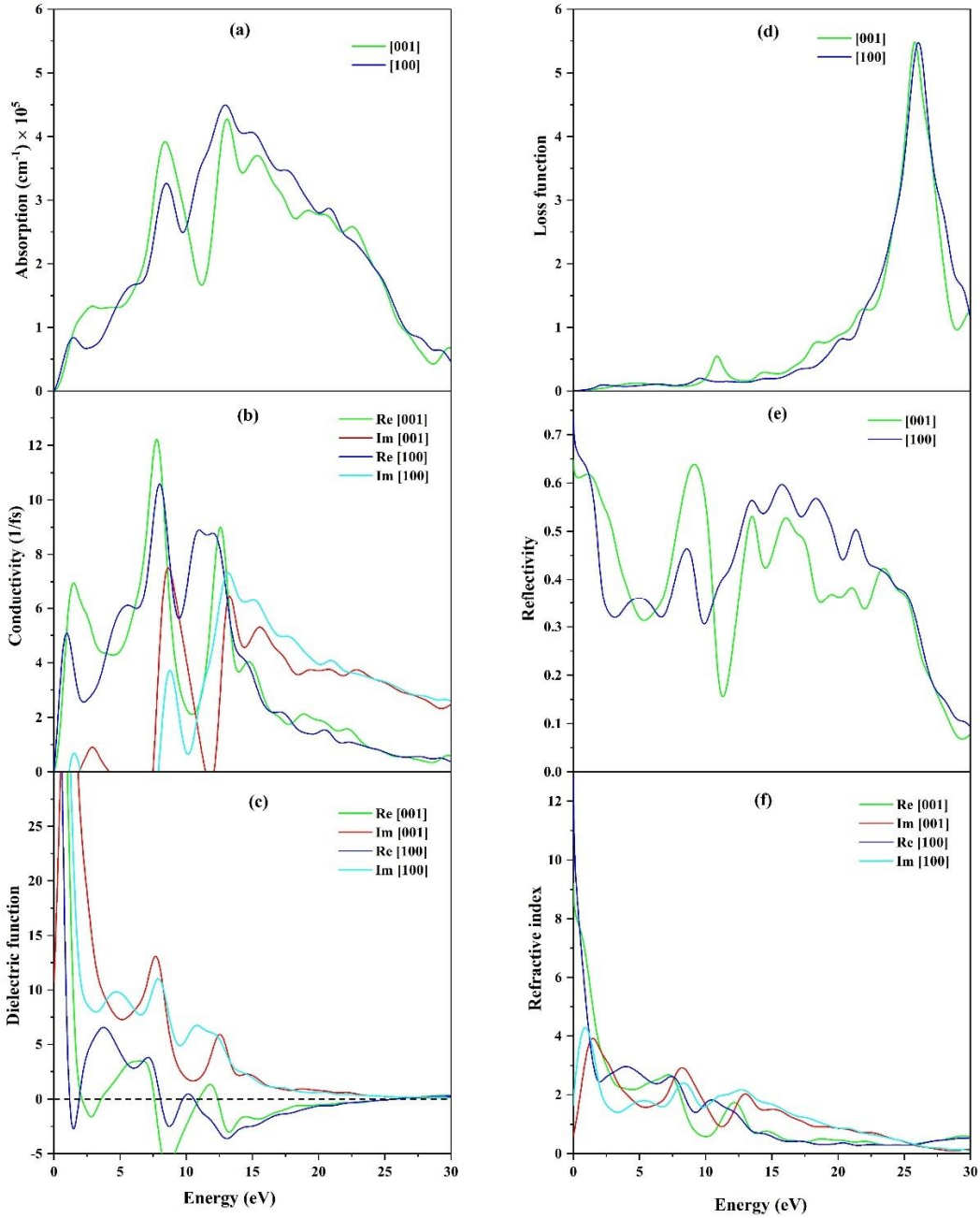

**Figure 8.** The energy dependent (a) absorption coefficient (b) optical conductivity (c) dielectric function (d) loss function (e) reflectivity, and (f) refractive index of CrB$_2$ with electric polarization vectors along the [100] and [001] directions.

Another crucial optoelectronic property of a material is its optical conductivity, $\sigma(\omega)$, which is the conduction of free charge carriers within a certain range of photon energy. An accurate predictor of photoconductivity is optical conductivity [106].

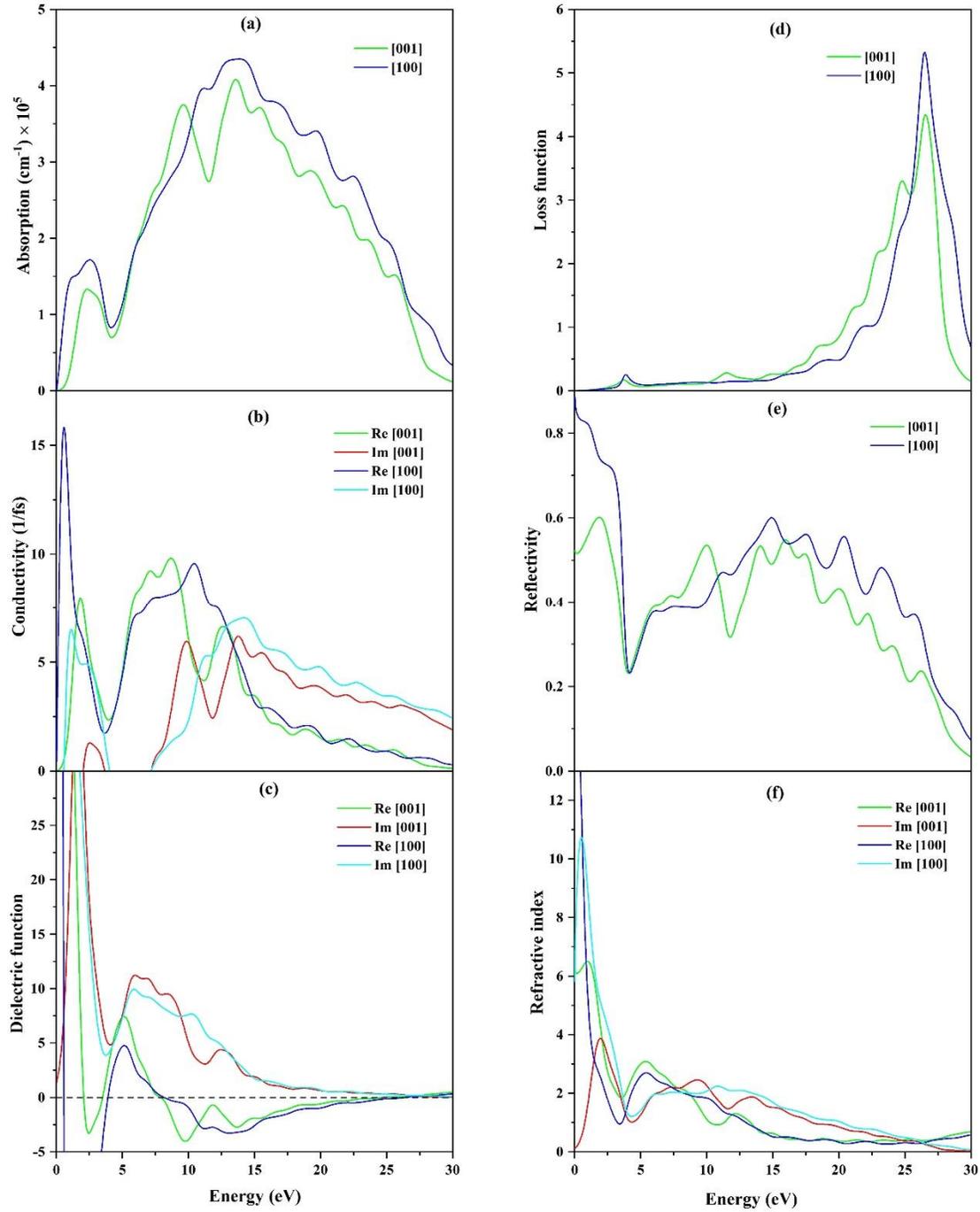

**Figure 9.** The energy dependent (a) absorption coefficient (b) optical conductivity (c) dielectric function (d) loss function (e) reflectivity, and (f) refractive index of $MoB_2$ with electric polarization vectors along the [100] and [001] directions.

Photoconductivity starts with zero photon energy in both polarization directions [Figures 8(b), 9(b), and 10(b)], which is a hallmark of metallic conductivity of $XB_2$ and is perfectly consistent with its electronic band structure and TDOS predictions. In the photon energy range of 0 to 15 eV for the compounds, the spectra for both polarizations show significant variation, indicating that their optical properties are anisotropic. In the mid-UV range, the optical conductivity falls with increasing energy.

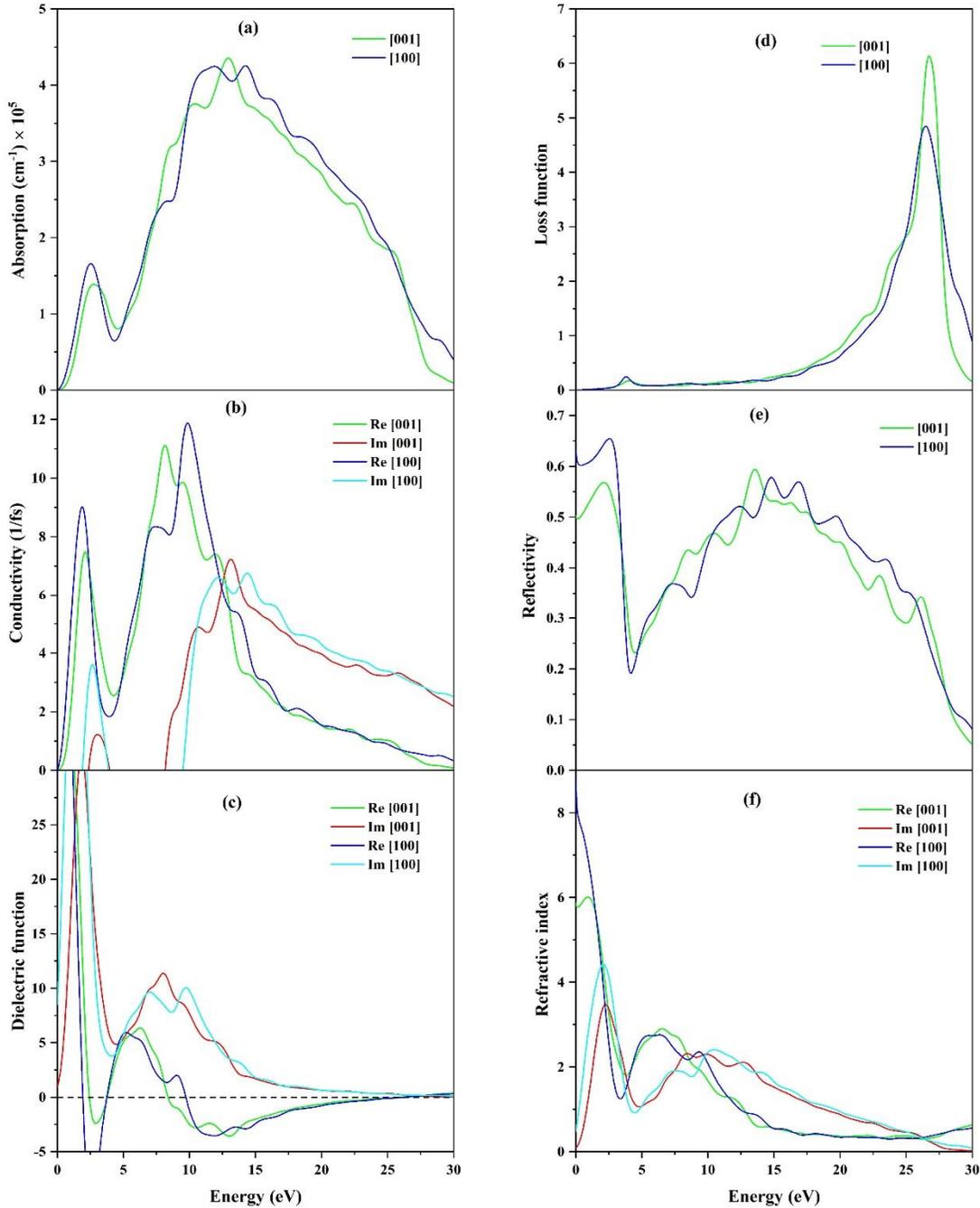

**Figure 10.** The energy dependent (a) absorption coefficient (b) optical conductivity (c) dielectric function (d) loss function (e) reflectivity, and (f) refractive index of $WB_2$ with electric polarization vectors along the [100] and [001] directions.

Figures 8(c), 9(c), and 10(c) display the real and imaginary parts of the dielectric constants. The real part $\varepsilon_1(\omega)$ is connected to the electrical polarization of the material, while the imaginary part $\varepsilon_2(\omega)$ is connected to the dielectric loss. The non-zero value of $\varepsilon_1(0)$ indicates the high availability of free charge carriers in these metallic systems. For the three compounds, there is a noticeable anisotropy in both spectra as a result of the differing polarizations. For all three compounds, the real part of the dielectric function crosses zero from below at ~ 26.50 eV, while the imaginary part flattens to a very low value at the same energy. This indicates that the material becomes transparent to the incident EMW above 26.50 eV. This phenomenon validates the Drude-like behavior (metallic feature) of $XB_2$. The frequency at which $\varepsilon_1(\omega)$ evolves from negative to positive values along with $\varepsilon_2(\omega) < 1$ is known as the plasma frequency, $\omega_P$. At this frequency, the free electrons in a system are collectively excited. For high frequency or power applications, low values of $\varepsilon_2(\omega)$ are favored in order to minimize electric power loss. Conversely, for small-sized capacitance applications, high values of $\varepsilon_1(\omega)$ are advised.

The loss function, $L(\omega)$, is an essential optical parameter describing the energy loss of a fast electron moving in a material. In this instance, the energy of the rapidly moving charge is attenuated, leading to a collective electronic excitation that is referred to as a plasma excitation. It is noteworthy that the plasma energies correspond with a rapid decrease in absorption coefficient and reflectance. As a result, it is anticipated that the compound under investigation would exhibit transparent behavior for photon energies higher than the plasma energy, and its optical characteristics will resemble those of insulating systems. Figures 8(d), 9(d), and 10(d) show that for [100] and [001] polarization directions, the peaks of $L(\omega)$ are located at ~ 26.50 eV. Above the plasma energy, optical behavior of a metal shifts to dielectric-like response. $MoB_2$ exhibits a higher peak value for [100] polarization direction than [001] polarization direction, $WB_2$ exhibits a higher peak value for [001] polarization direction than [100] polarization direction, and $CrB_2$ exhibits a slightly higher peak value for [001] polarization direction, indicating optical anisotropy in the loss function.

Reflectivity, $R(\omega)$, measures the fraction of the incident light energy reflected from the material. The compounds $XB_2$ (X = Cr, Mo, and W) exhibit excellent infrared and visible light reflectivity, as seen by Figures 8(e), 9(e), and 10(e). The highest reflectivity values in the infrared light region due to polarization [100] is 68.70% for $CrB_2$, 85.60% for $MoB_2$, 61.60% for $WB_2$ while for polarization [001] the corresponding values are 61.90%, 52.40%, and 49.50%, respectively. For both polarizations, the average reflectivity of $XB_2$ within visible light region is above 44%, which is enough for them to be a potential coating material to diminish solar heat [38,103]. It is observed that the maximum average reflectivity is found for $MoB_2$.

The energy dependent refractive index, $N(\omega)$, is an optical parameter important for photonic device fabrication, such as for constructing optical wave-guides. In fact, $N(\omega) = n(\omega) + ik(\omega)$, where $k(\omega)$ is the extinction coefficient. The refractive index $n(\omega)$ (real part of complex index of refraction) of a material is a number that refers to how fast light travels through the material.

The amount of light that is reflected at the contact is likewise determined by the refractive index. Throughout the visible spectrum, the refractive index of the majority of materials changes by a few percent with photon energy. The $n(\omega)$ spectra (Figures 8(f), 9(f), and 10(f)) for all the three compounds exhibit anisotropic nature. Compared to $CrB_2$ and $WB_2$, the average real part of refractive index of $MoB_2$ has a very high value at low energies encompassing the infrared regions. Systems with a high refractive index provide advantageous optical characteristics for optical display devices and photonic crystals.

The imaginary component of the complex index of refraction, known as the extinction coefficient, regulates how quickly light is occluded. In addition, it quantifies the strength of a material's absorption of light at a certain wavelength per mass density or per molar concentration and characterizes the decrease of electromagnetic radiation in a medium. The extinction coefficient is associated with the conductive properties of materials. The extinction coefficient of metallic materials is significant, whereas that of semiconducting materials is minimal. In contrast, dielectric materials are essentially non-conductors whose extinction coefficient is zero. A large static value of $k(\omega)$ of $MoB_2$ indicates that the metallic conductivity of $MoB_2$ is greater than that of $CrB_2$ and $WB_2$. In the high energy region where $n(\omega)$ begins to increase and achieves an almost constant value for all compounds, $k(\omega)$ becomes zero.

### 3.6 Superconducting state properties

$XB_2$ (X = Cr, Mo and W) compounds exhibit phonon mediated superconductivity. For conventional superconductors where electron–phonon interactions lead to Cooper pairing, the superconducting transition temperature can be expressed as [38,85,107,108]:

$$T_c = \frac{\theta_D}{1.45} exp\left[-\frac{1.04(1 + \lambda_{ep})}{\lambda_{ep} - \mu^*(1 + 0.62\lambda_{ep})}\right] \qquad (53)$$

where $\mu^*$ is the Coulomb pseudopotential and $\lambda_{ep}$ is the electron-phonon coupling constant.

The electron-electron interaction parameter of a material, termed as the repulsive Coulomb pseudopotential, can be calculated as follows from the electronic density of states at the Fermi energy [110]:

$$\mu^* = \frac{0.26\, N(E_F)}{1 + N(E_F)} \qquad (54)$$

where, $N(E_F)$ is the total density of states at the Fermi level of the compound. The repulsive Coulomb pseudopotential is responsible to lower the transition temperature, $T_c$, of superconducting compounds [109,110]. It also gauges how strongly electrons are correlated inside a system. The electron-phonon coupling constant can be estimated approximately using the equation presented below [111,112]:

$$\lambda_{ep} = \frac{\gamma_{cal}}{\gamma_{free}} - 1 \qquad (55)$$

where $\gamma_{cal}$ is the theoretical electronic specific heat capacity constant calculated from the DOS at Fermi level and $\gamma_{free}$ is the theoretical electronic specific heat capacity constant estimated from the free electron gas model. In this work, we have calculated $\gamma_{cal}$ and $\gamma_{free}$ of $XB_2$ materials by using the following equations [112-114]:

$$\gamma_{cal} = \frac{\pi^2 k_B^2 N(E_F)}{3}$$

$$\gamma_{free} = \frac{\pi^2 n_e k_B^2}{2E_F}$$

where, $n_e$ denotes the free electron number density and $E_F$ is the Fermi energy. The resulting $\gamma_{cal}$ value obtained using the $N(E_F)$ from the band structure calculations differs from the $\gamma_{free}$ of $XB_2$. This is expected since the DOS value obtained from the free electron model is not *dressed* properly by the electron-ion interaction.

Table 12 displays the determined superconducting parameters, including the critical temperatures of $XB_2$ compounds. The $T_c$ values obtained show reasonable agreement with the experimental ones.

Table 12. Parameters related to superconductivity of $XB_2$ (X = Cr, Mo and W) compounds.

| Compound | $N(E_F)$ (States per eV) | $\mu^*$ | $\gamma$ (mJ/mol-K$^2$) | $\lambda_{ep}$ | $T_c$ (K) | Reference |
|---|---|---|---|---|---|---|
| CrB$_2$ | 2.52 | 0.19 | 5.94 | 0.67 | 8.09 | This work |
|  | - | - | - |  | 7.00 | [21] |
| MoB$_2$ | 1.34 | 0.15 | 3.16 | 0.88 | 27.12 | This work |
|  | - | 0.13 | - | 1.67 | 32.00 | [19] |
| WB$_2$ | 1.25 | 0.14 | 2.95 | 0.57 | 6.02 | This work |
|  | - | - | - | - | 5.40 | [115] |

## 4. Conclusions

To summarize, we have presented a comprehensive study of the structural, mechanical, electronic, optical, superconducting state and thermophysical properties of binary metal diborides $XB_2$ (X = Cr, Mo and W) using a first-principles calculations based on density functional theory. Our calculated equilibrium structural parameters and elastic constants of $XB_2$ are consistent with the previously published data [11-13,15,22].

XB$_2$ (X = Cr, Mo and W) compounds have high machinability, high hardness values, and an exceptionally high melting temperature, making them ideal for use as heavy-duty engineering tools that can withstand highly hostile and abrasive environments. Importantly, WB$_2$ is mechanically superior to MoB$_2$ and CrB$_2$, because its Young's modulus, shear modulus and hardness are higher than those of MoB$_2$ and CrB$_2$. WB$_2$ is also ductile in nature, whereas MoB$_2$ and CrB$_2$ are brittle. Ductility together with very high hardness makes WB$_2$ a very useful compound for structural engineering. The band structure and total density of states investigations indicate that XB$_2$ (X = Cr, Mo and W) compounds are metallic with considerable TDOS at the Fermi level. For all of the XB$_2$ (X = Cr, Mo, and W) compounds, the computed Debye temperatures are high, suggesting their hard nature. High melting points and lattice thermal conductivity values match extremely well to estimated hardness and elastic moduli. The optical parameters exhibit metallic behavior consistent with the electronic band structure. The anisotropy levels in elastic and optical properties of WB$_2$ are less than those of MoB$_2$ and CrB$_2$. All of the compounds investigated here be utilized as a coating to minimize solar heating, with MoB$_2$ being the most promising contender. All the compounds under study are effective UV radiation absorbers. We have explored the superconducting state properties of XB$_2$ (X = Cr, Mo and W) and found that the compound MoB$_2$ shows the highest superconducting transition temperature. The high-T$_c$ results from a high electron-phonon coupling constant for MoB$_2$ compared to the other two metallic diborides.

## Declaration of interest

The authors declare that they have no known competing financial interests or personal relationships that could have appeared to influence the work reported in this paper.

## Data availability

The data sets generated and/or analyzed in this study are available from the corresponding author on reasonable request.

## CRediT author statement

**Razu Ahmed:** Methodology, Software, Formal analysis, Writing- Original draft. **Md. Sohel Rana:** Methodology, Software, Formal analysis. **Md. Sajidul Islam**: Software, Validation. **S.H. Naqib:** Conceptualization, Supervision, Formal analysis, Writing- Reviewing and Editing.